\documentclass[12pt]{article}

\usepackage{scicite}

\usepackage{times}
\usepackage{authblk}
\usepackage{setspace}
\usepackage{upgreek}
\usepackage{graphicx}
\usepackage{mathtools}
\usepackage{hyperref}
\usepackage{fixltx2e}
\usepackage{xcolor}
\usepackage{epstopdf} 
\usepackage[normalem]{ulem}
\usepackage{caption}
\usepackage{url}
\usepackage{color}
\usepackage{tocloft}
\usepackage{amsmath}
\usepackage{mathrsfs}
\usepackage[normalem]{ulem}
\usepackage{caption}
\usepackage{graphicx}

\newenvironment{sciabstract}{%
\begin{quote} \bf}
{\end{quote}}

\newcounter{lastnote}

\title{Chip-Scale Continuously Tunable Optical Orbital Angular Momentum Generator}

\author
{\small Jie Sun,$^{1}$ Ami Yaacobi,$^{1}$ Michele Moresco,$^{1}$ Douglas Coolbaugh,$^{2}$ Michael R. Watts$^{1\ast}$\\
\normalsize{\vspace{3.0mm}$^{1}$Research Laboratory of Electronics, Massachusetts Institute of Technology, \\Cambridge, Massachusetts 02139, USA}\\
\normalsize{\vspace{3.0mm}$^{2}$College of Nanoscale Science \& Engineering, University at Albany, \\State University of New York, Albany, New York 12203, USA}\\
\normalsize{\vspace{8.0mm}$^\ast$To whom correspondence should be addressed; E-mail: {\color{blue}\underline{\url{mwatts@mit.edu}}}.}
}

\date{}

\begin{document} 


\baselineskip24pt


\maketitle


\begin{sciabstract}
Light carrying orbital angular momentum (OAM) has potential to impact a wide variety of applications ranging from optical communications to quantum information and optical forces for the excitation and manipulation of atoms, molecules, and micro-particles. The unique advantage of utilizing OAM in these applications relies, to a large extent, on the use of multiple different OAM states. Therefore, it is desirable to have a device that is able to generate light with freely adjustable OAM states in an integrated form for large-scale integration. We propose and demonstrate a compact silicon photonic integrated circuit to generate a free-space optical beam with OAM state continuously tuned from a single electrical input signal, realizing both integer and non-integer OAM states. The compactness and flexibility of the device and its compatibility with complementary metal-oxide-semiconductor (CMOS) processing hold promise for integration with other silicon photonic components for wide-ranging applications.  
\end{sciabstract}


\section*{}

\paragraph*{One Sentence Summary:} We demonstrated a compact silicon photonic integrated circuit to generate light with continuously tunable orbital angular momentum. 

\section*{}
\paragraph*{Main Text:} An optical vortex beam, where the electromagnetic field has an azimuthal phase dependence ($E(\theta)\sim e^{i\cdot l\cdot\theta}$  where $\theta$ is the azimuthal angle), possesses a helical phase front, and therefore carries an orbital angular momentum (OAM) $L=l\cdot\hbar$ per photon where $\hbar$ is the reduced Planck constant \cite{ref1,ref2}. The $l$-value represents the OAM state, which is usually but not necessarily an integer \cite{ref3,ref4,ref5}, and its sign indicates the chirality or handedness of the OAM with respect to the beam direction, either clockwise or counterclockwise. Optical vortex beams have found a wide range of applications including optical communications \cite{ref6,ref7}, quantum information \cite{ref8,ref9,ref11,ref12}, and optical micromanipulation \cite{ref13,ref14,ref15}, to name just a few. Although it is challenging to accurately engineer the sensitive optical phase to generate well-defined OAM beams, a variety of approaches have been successfully demonstrated, including helically twisted photonic crystal fibers \cite{ref16}, free-space optical components such as liquid crystal based spatial light modulators \cite{ref6,ref7}, spiral phase plates \cite{ref5,ref11}, and recently designer metasurfaces \cite{ref17} using surface plasmonics which has shown precise and broadband engineering \cite{ref18} of free-space wavefronts.  

In addition, silicon photonic integrated circuits have emerged as a new tool for optical wavefront synthesis and shown remarkable accuracy in controlling the optical phase at large scale \cite{ref19}. Silicon photonics is therefore also useful for generating OAM beams, with advantages in compactness, active tunability, and the promise for large-scale on-chip integration with other silicon photonic devices and functionalities. Indeed, several photonic integrated circuits have been demonstrated to generate OAM beams, including a relatively large-size photonic circuit with complex electrical controls \cite{ref20,ref21} and a very compact device based on microring resonators \cite{ref22,ref23}. However, for applications such as the space-division multiplexing (SDM) in optical communications, multiple and reconfigurable OAM states at a given working wavelength provide an additional multiplexing dimension to further increase the data capacity \cite{ref6,ref7}.  Moreover, for applications such as quantum information processing, non-integer OAM states, which are essentially a superposition of a number of integer OAM states, can be particularly useful in, for example, conveniently detecting high-dimensional entanglement \cite{ref11,ref12}.  In addition, for applications such as optical manipulation where the OAM of the photons directly transfers to that of a trapped particle causing it to rotate \cite{ref15}, the capability to continuously adjust the OAM state to both integer and non-integer $l$-values offers fine control and enhanced functionalities in optical manipulation \cite{ref24,ref25}. As a result, it is desirable to have a device to generate continuously tunable OAM states, preferably in an integrated form to provide chip-scale solutions for these applications. To this end, silicon photonic circuits have the unique advantage for its capability to achieve efficient active phase tuning, a necessary feature, for implementing tunable OAM states. While prior demonstrations have produced different OAM states by coupling to different ports on an otherwise passive chip \cite{ref20} or by adjusting the input wavelength \cite{ref23}, no freely tunable OAM generator has previously been demonstrated.  Here we propose and demonstrate a compact silicon photonic circuit to generate optical vortex beams with well-defined OAM; in particular, we demonstrate that, by applying a single electrical signal, the generated OAM beam can be continuously and wavelength-independently tuned over a broad wavelength range to achieve not only the conventional integer OAM states but also non-integer states, with both clockwise and counterclockwise chirality of the emitted phase front. 

A schematic of the proposed OAM generator is shown in Fig. 1A. Light is launched from an optical fiber to a silicon bus waveguide that forms an open circle. Light in the bus waveguide evanescently couples into $N$ unit cells in sequence as it travels clockwise along the loop. Figure 1B shows a close-up view of the unit cell, consisting of a directional coupler, an optical emitter, and a thermo-optically tunable phase shifter. The directional coupler drops optical power from the bus waveguide into the unit cell in such a manner that each unit cell receives the same amount of power, achieved by varying the coupling efficiency controlled by the gap and length of the directional couplers. Light in the unit cells is then routed to and eventually emits from the grating-based optical emitters that are closely packed in a circle in the center of the loop (Fig. 1A). The emitter adopts the design in \cite{ref19}, which features a compact size (3.0$\upmu$m$\times$2.8$\upmu$m), high up-emitting efficiency (51\%), and broadband emission ($>$200nm). The adjacent unit cells are connected by a thermo-optically tunable phase shifter which generates an electrically controlled tunable phase shift $-\Delta\phi$ between two unit cells. As shown in Fig. 1B, the embedded tunable phase shifter is realized by doping the silicon waveguide to form a resistive heater, which provides for a highly efficient thermo-optic phase shifter in a compact footprint that fits within the unit cell \cite{ref26}. Adiabatic waveguide bends are used here to bring in silicon electrical contacts without introducing much scattering loss \cite{ref19,ref26}, an important feature since the phase shifters are cascaded in series and the optical loss is accumulated as the light travels along the bus waveguide loop. These identical phase shifters are electrically connected in parallel (Fig. 1A) in order to generate the same phase delay $-\Delta\phi$ when a tuning voltage is applied. In addition, the routing waveguides in individual unit cells are passively phase matched so that the relative phase of the optical emission in adjacent unit cells depends only on the phase delay $-\Delta\phi$. Therefore, the $m^{th}$ optical emitter along the light propagation direction (clockwise) has an emitted phase $-m\cdot\Delta\phi$. The resulting optical field, formed by the optical emission from all of the $N$ optical emitters, is then given by $\vec{E}(\theta)\sim\vec{t}\cdot e^{-i\cdot m\cdot\Delta\phi}$, where $\vec{t}$ is the direction of the electric field of the optical emission from each emitter. Since the unit cells are equally separated in the azimuthal direction by an angle $\Delta\theta=2\pi/N$ (Fig. 1A), the $m^{th}$ emitter has an azimuthal angle $\theta=-m\cdot\Delta\theta$, where the minus sign results from the counterclockwise convention of the azimuthal angle. If we rewrite $\Delta\phi$ in terms of $\Delta\theta$
\begin{equation}
\Delta\phi=l\cdot\Delta\theta
\end{equation}
we can immediately find the optical field can be written as $\vec{E}(\theta)\sim\vec{t}\cdot e^{i\cdot l\cdot\theta}$, which resembles the phase profile of an optical vortex beam possessing an OAM state $l$. Since the phase shift $\Delta\phi$ is electrically controlled by the voltage applied on the phase shifters, the $l$-value can be continuously tuned accordingly. Compared to the optical vortex beam emitter with a resonant cavity \cite{ref22,ref23}, the non-resonant device proposed here eliminates the wavelength dependence and hence enables continuously tunable OAM states to be generated at any given wavelength with both integer and non-integer $l$-values. 

The proposed device was fabricated in a CMOS-compatible silicon photonic process on a silicon-on-insulator (SOI) wafer with 0.22-$\upmu$m thick top silicon layer to form the waveguides (Fig. 1C) and emitters (Fig. 1D). Two different levels of dopings, $n$ and $n^+$, were applied to create the low-loss integrated thermo-optic phase shifter. The low-concentration $n$-doping was used in the silicon waveguide to minimize the scattering loss caused by the dopants as well as to increase the resistance so that most heating power drops on the waveguide. And the high-concentration $n^+$-doping was utilized for the low-resistance silicon contacts (Fig. 1B and Fig. 1E). 

A laser input at 1493 nm was coupled into the transverse-electric (TE) mode of the silicon bus waveguide to excite all of the emitters, forming a circular ``necklace'' pattern with a radius $R=17 \upmu$m in the near field (Fig. 2A). Thirty emitters were chosen ($N=30$) in this work.  As shown in Fig. 2A, uniform emission was observed across all thirty optical emitters. Considering that the thermo-optic phase shifters are cascaded in the bus waveguide and the optical emission from the last unit cell in the loop experiences an accumulated loss from all of the previous 29 phase shifters, the uniformity highlights the ultra-low insertion loss of the integrated phase shifter as well as the accuracy of the directional coupler design and the silicon photonic fabrication. Therefore, it is straightforward to incorporate more unit cells to increase the angular resolution of the OAM generator. 

Since the photonic circuit operates predominantly in the TE polarization on-chip, the optical emission from each emitter is azimuthally polarized with the electric field pointing tangentially to the circle centered on the beam axis (Fig. 1B), essentially a cylindrical vector vortex beam with many potential applications \cite{ref27}. Although the optical emission of each emitter is linearly polarized with respect to itself, the state of polarization of the synthesized field as a whole is intricate. However, this complex polarization state can be seen as a superposition of ordinary polarizations, as discussed in the following. For an optical emitter at azimuthal angle $\theta$ (Fig. 1B), its electric field can be decomposed into $x$- and $y$-direction, described by the Jones vector \cite{ref28}
\begin{equation}
\vec{E}\sim\begin{pmatrix} -\sin\theta \\ \cos\theta \end{pmatrix}\cdot e^{i\cdot l\cdot\theta}=\frac{i}{2}\begin{pmatrix} 1\\ -i \end{pmatrix}\cdot e^{i\cdot (l+1)\cdot\theta}-\frac{i}{2}\begin{pmatrix} 1\\ i \end{pmatrix}\cdot e^{i\cdot (l-1)\cdot\theta}
\end{equation}
which means that the generated optical emission is a combination of a right-hand circularly polarized (RHCP) vortex beam with OAM state $l_R=l+1$ and a left-hand circularly polarized (LHCP) vortex beam with OAM state $l_L=l-1$. It is also noticed that, since the RHCP beam carries a spin orbital momentum $-\hbar$ and the LHCP beam carries $\hbar$, the overall angular momentum of these two polarizations, including the orbital angular momentum and the spin angular momentum, is exactly the same, which is $l\cdot\hbar$, meaning the total angular momentum is conserved regardless of the polarization state. An experimental demonstration of the polarization decomposition is shown in Fig. 2B. An $l=+1$ vortex beam was first synthesized at the working wavelength by adjusting the voltage applied on the tunable phase shifter (left column, Fig.2B). With a quarter-wave plate, the RHCP and LHCP components in the synthesized field were projected into linear polarized light on the axes 45$^{\circ}$ and 135$^{\circ}$ to the fast axis of the quarter-wave plate, respectively, as shown by Fig. S1 \cite{refsp}. The RHCP component in the synthesized field was then revealed (center column of Fig. 2B) by filtering out the LHCP component with an additional linear polarizer. The RHCP beam (OAM state $l_R=+2$) has an annular beam profile with a phase singularity in the center where the phase is undefined, a signature of the high-order vortex beams. Likewise, the LHCP beam ($l_L=0$) was also imaged (right column of Fig. 2B), showing a focused bright spot in the center where constructive interference occurs when $l_L=0$.

Another method to visualize the helical phase front of the vortex beam is to interfere with a co-propagating Gaussian reference beam to create a spiral interference pattern from which the $l$-value of the OAM state can be distinguished by the number of the arms in the spiral while the sign of the OAM is revealed by the chirality of the spirals \cite{ref2,refsp}. As shown in Fig. 3A, the signature spiral pattern has been observed by interfering the RHCP component of the synthesized field with a co-propagating RHCP Gaussian beam using the setup illustrated in Fig. S2. From Fig. 3, we see that that well-defined OAM states have been generated by the proposed device. By applying voltages on the cascaded heaters to tune the phase shift $\Delta\phi$, according to Eq. 1, the OAM state ($l$-value) can thus be dynamically reconfigured in a continuous way at the given wavelength, as shown in Fig. 3A. Again, measurements agree well with simulations, highlighting the accuracy of silicon photonic circuit in engineering the optical phase and the robustness of silicon photonic fabrication. As shown in Fig. 3B, the OAM state varies continuously from -4 to 4 at the working wavelength by adjusting the heater power linearly from 3.5 mW to 10 mW. The corresponding phase shift is $\Delta\phi=\Delta l\cdot\Delta\theta=0.27\pi$, and the thermo-optic efficiency of the fabricated heater is hence 12.2 mW per $\pi$ phase shift. The high thermo-optic efficiency is a result of the direct integration of the heater in the waveguide that provides the largest overlap of the optical mode with the generated heat. According to the response time of a similar thermo-optic heater \cite{ref26}, the estimated reconfiguration time of the tunable OAM generator is on the order of a microsecond, much faster than that of the liquid-crystal based spatial light modulator which is usually on the order of a millisecond.  Moreover, in addition to the integer OAM states that have also been realized by previous demonstrations, the proposed OAM device was able to generate non-integer OAM states as well, which reveals a continuum of states inbetween two consecutive integer OAM states and dynamically shows how the vortex beam evolves from one integer state to the next (Movie S1 and S2). For example, the OAM state $l_R=+1.5$ in Fig. 3A shows the transition from +1 OAM state to +2 OAM state, where a dislocation emerges in the spiral arm, gradually moves to the center and finally develops into a separate spiral arm as $l_R$ approaches +2. The ability to continuously tune the vortex beam to both integer and non-integer OAM states comes from the non-resonant structure of the device which does not rely on a resonant cavity mode that would otherwise prevent non-integer OAM states from being constructed \cite{ref23}. The non-interger OAM states can be particularly useful in, for example, quantum information processing to conveniently produce and detect high-dimensional quantum entanglement as well as in optical manipulations \cite{ref11,ref12} to fine control the rotation of a trapped particle \cite{ref24,ref25}. 

The non-resonant structure of the proposed device also allows for broadband operation. For example, the device generated OAM state $l_R=-3$ at four different wavelengths 1466 nm, 1490 nm, 1515 nm, and 1541 nm when the heater was off (Fig. 4A), while the OAM state was tuned to $l_R=0$ at these wavelengths when 2.8 mW heating power was applied on each heater (Fig. 4B). This enables multiple wavelengths to be simultaneously projected to the same and tunable OAM state using a single device and a single electrical control, which is particularly useful in SDM optical communications where each OAM state is shared by multiple wavelengths to incorporate wavelength-division multiplexing (WDM) at the same time \cite{ref6,ref7}. The wavelength spacing is determined by the optical path length between two unit cells and can be adjusted accordingly to match the WDM wavelength standard. 

Our device, a tunable OAM generator and through reciprocity a tunable OAM detector, enables continuous and wavelength-independent tuning of the OAM state over a broad wavelength range and thus represents a leap forward in the development OAM technologies. Already, OAM is impacting a wide range of applications, and while much greater opportunities lie in the promise to integrate our device with other silicon photonic components and functionalities as well as electronics to realize chip-scale tunable OAM systems, with the aide of large-scale CMOS-compatible silicon photonic integration \cite{ref19}. By integrating with lasers, high-speed silicon modulators and detectors, compact SDM-WDM transceivers can be expected. And, with the recently developed CMOS-compatible single-photon detectors \cite{ref29}, high-dimensional quantum information processing can be realized on a silicon chip. Further, integration with microfluidics will enable on-chip optical manipulations such as tunable optical spanners, and integration with on-chip optical interferometers will provide chip-scale solution for optical sensing such as detecting the spin of an object \cite{ref30}. Finally, the device opens up new possibilities to generate complex and reconfigurable optical beamforms including even vector beams using chip-size, large-scale silicon photonic circuits.

\paragraph{Acknowledgement:} We thank X. Cai and S. Yu (University of Bristol, UK) for helpful discussions, and D. B. Cole and J. Zhou (Massachusetts Institute of Technology, USA) for the help with the measurement setup. This work was supported by the Defense Advanced Research Projects Agency (DARPA) of the United States under the E-PHI project, grant no. HR0011-12-2-0007.

\clearpage
\begin{figure}
\centerline{\includegraphics [width = 15.3cm] {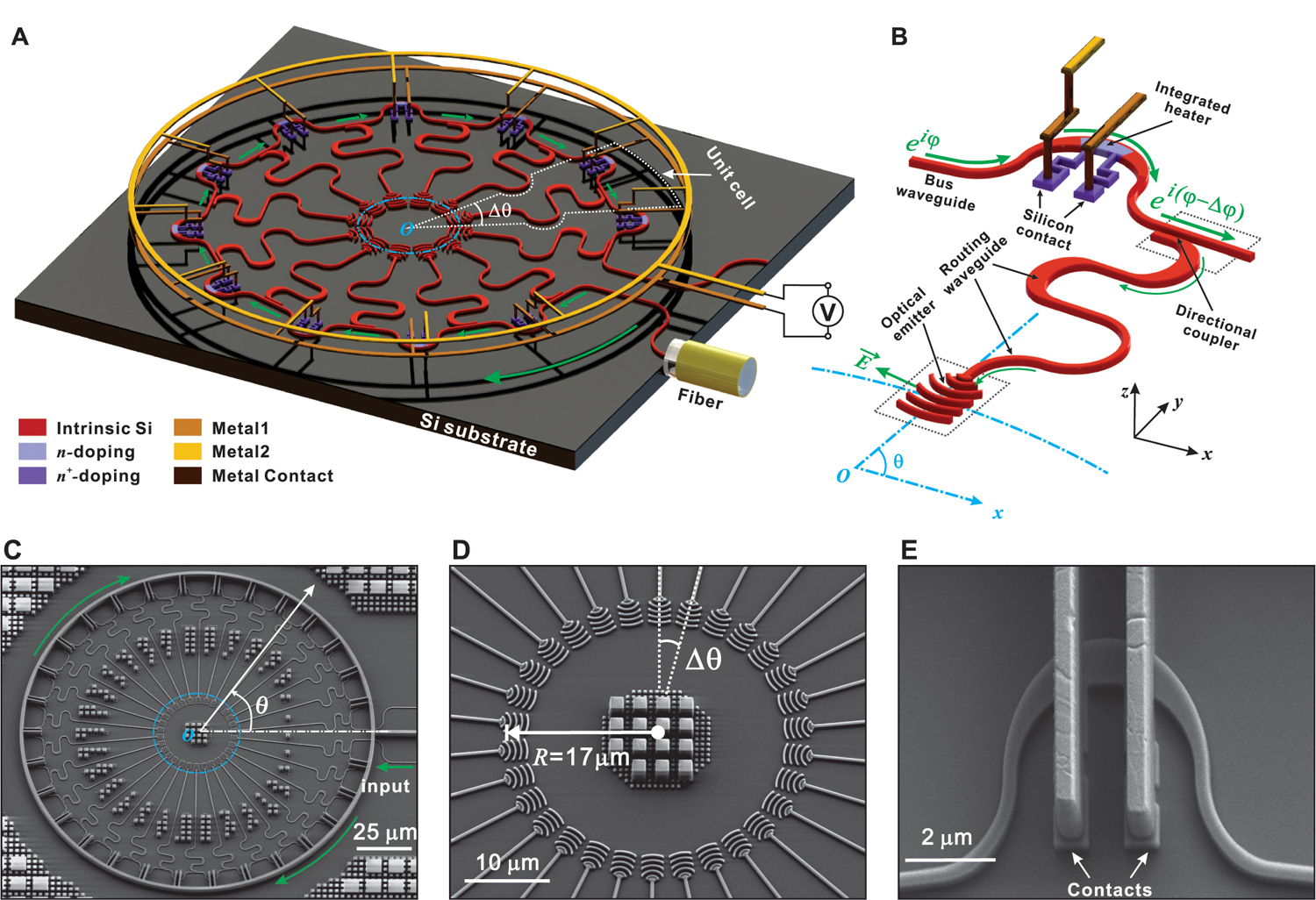}}
\end{figure}
\noindent {\bf Fig. 1.} \textbf{The structure of the continuously tunable OAM generator.} \textbf{(A)} A schematic of the proposed continuously tunable integrated OAM generator, consisting of $N$ unit cells placed concentrically along a silicon bus waveguide that forms an open circle. \textbf{(B)} A schematic of an individual unit cell, consisting of a directional coupler, a directly integrated thermo-optic phase shifter, and a grating-based optical emitter. Scanning-electron micrographs (SEMs) of \textbf{(C)} the fabricated OAM generator using a CMOS-compatible silicon photonic process, \textbf{(D)} a close-up view of $N=30$ optical emitters in the center of the device, and \textbf{(E)} a close-up view of the integrated thermo-optic phase shifter. The SEMs were taken by dry-etching away the silicon-dioxide overcladding to reveal all of the metal layers and the silicon waveguides. The scattered squares inbetween the waveguides are metal and silicon fill structures to ensure the pattern density for the subsequent chemical-mechanical polishing steps in the process, but have no other optical or electrical purpose.  

\begin{figure}
\centerline{\includegraphics [width = 16.0cm] {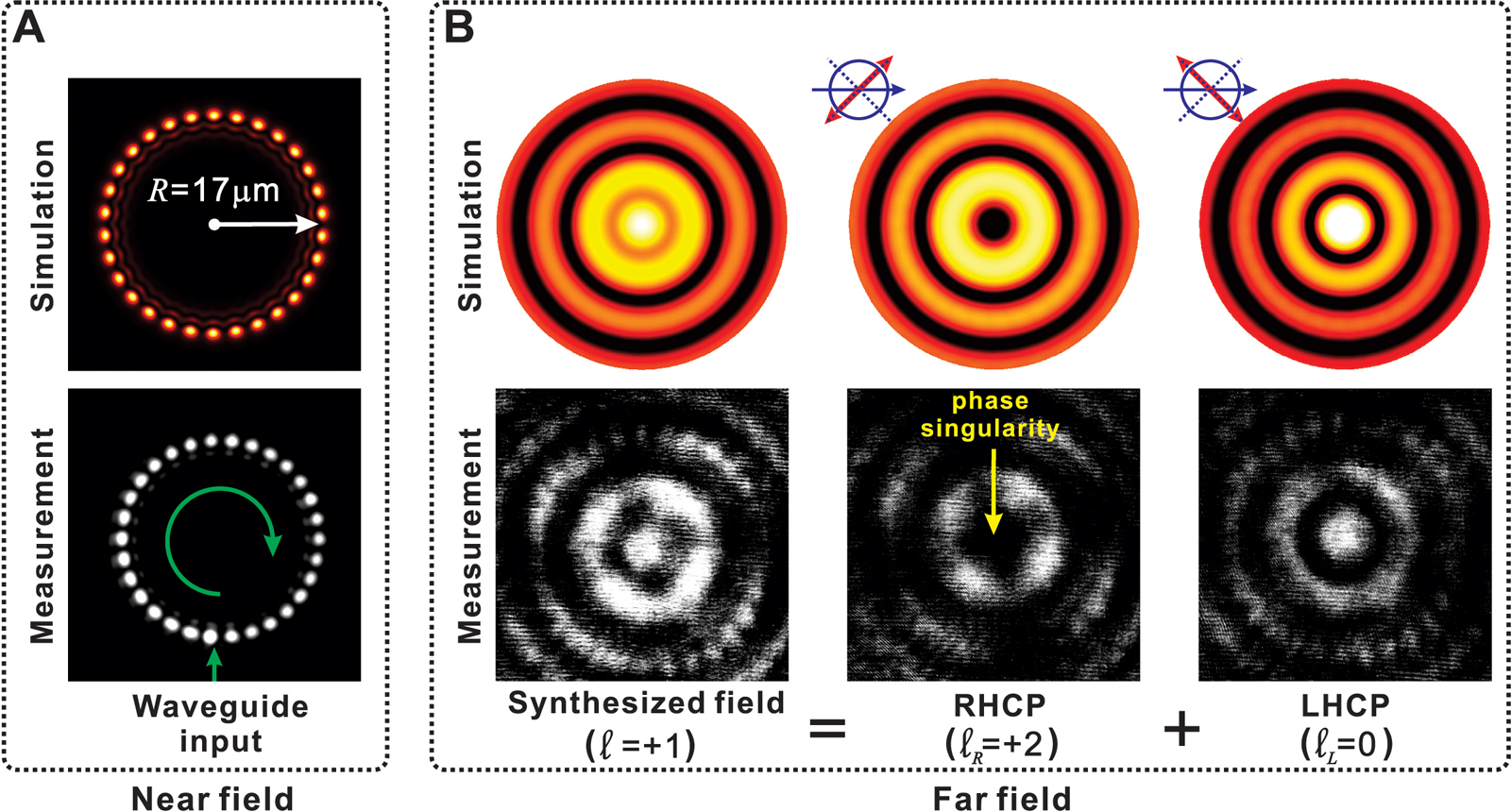}}
\end{figure}
\noindent {\bf Fig. 2.} \textbf{Near-field and far-field characterizations.} \textbf{(A)} The simulated (top) and measured (bottom) near-field emission of the device. Uniform emission across all of the thirty optical emitters has been achieved. The green circle represents the propagation direction of light. \textbf{(B)} Demonstration of the polarization decomposition of the synthesized vector beam. The far field of the vector vortex beam generated by the device (left, OAM state $l$) can be decomposed into an RHCP vortex beam (center, OAM state $l_R=l+1$) and an LHCP vortex beam (right, OAM state $l_L=l-1$), with balanced intensity. The LHCP (RHCP) beam was observed by filtering out the RHCP (LHCP) component with a quarter-wave plate and a linear polarizer. The dark-intensity region in the center of the RHCP vortex beam ($l_R=+2$) corresponds to a phase singularity since the optical phase is undefined in the center. Measurements (bottom row) agree well with the simulations (top row). 

\clearpage
\newpage
\begin{figure}
\centerline{\includegraphics [width = 15.9cm] {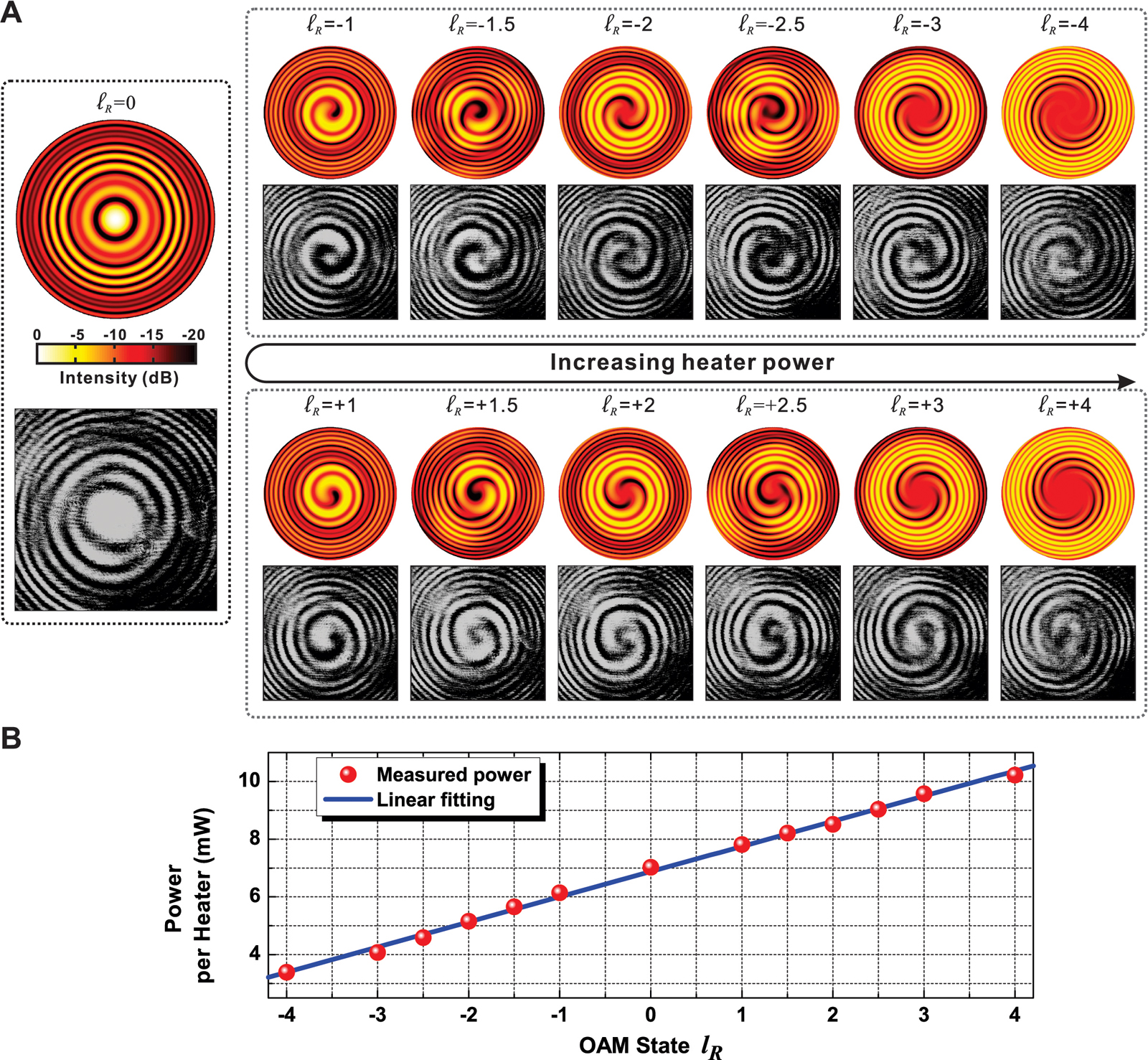}}
\end{figure}
\noindent {\bf Fig. 3.} \textbf{Demonstration of the continuously tunable OAM state.} \textbf{(A)} The OAM state of the generated vortex beam can be electrically tuned by the voltage applied on the thermo-optic phase shifter. OAM states from -4 to +4 have been demonstrated, revealed by the number of arms and the chirality of the spiral interference pattern with a co-propagating RHCP Gaussian beam. The non-integer OAM states ($\pm1.5$, $\pm2.5$) have also been demonstrated, showing the beam evolution inbetween two integer OAM states. A good agreement between simulation (top row) and measurement (bottom row) has been observed. \textbf{(B)} The measured power applied on the phase shifter to achieve different OAM states, which increases linearly as the OAM state is continuously tuned from -4 to +4. 

\clearpage
\newpage
\begin{figure}
\centerline{\includegraphics [width = 15cm] {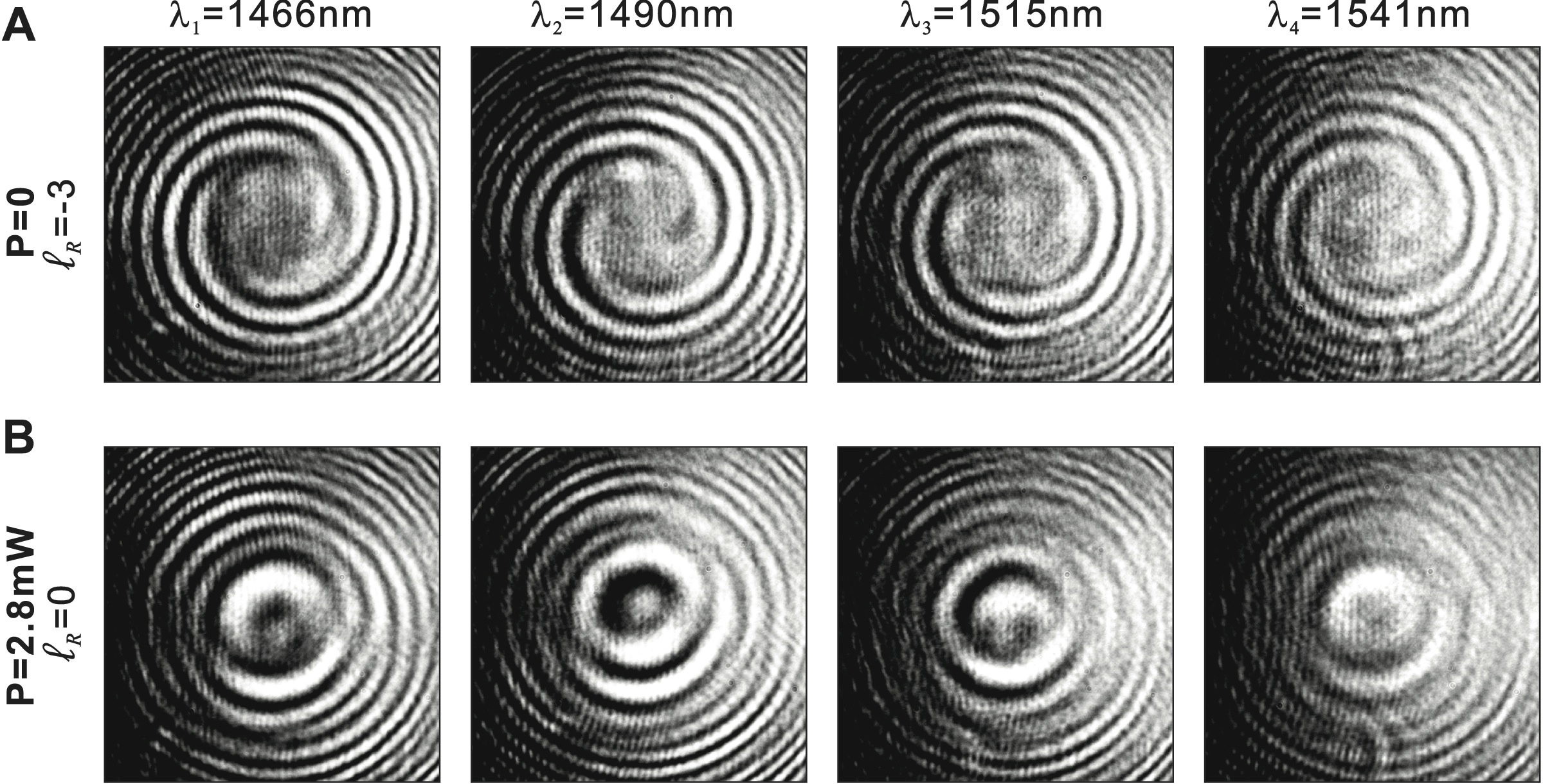}}
\end{figure}
\noindent {\bf Fig. 4.} \textbf{Broadband operation of the OAM generator.} \textbf{(A)} When the heater was off, OAM state $l_R=-3$ was generated at four different wavelengths, 1466 nm, 1490 nm, 1515 nm, and 1541 nm. \textbf{(B)} When 2.8 mW power was applied on each heater, the OAM state was switched to $l_R=0$ at these wavelengths. This can be used in SDM-WDM optical communications to simultaneously project multiple wavelengths into the same and tunable OAM state using a single device with simple electrical control.

\clearpage
\newpage
\section*{\huge{Supplementary Information}}
\vspace{16pt}

\renewcommand\contentsname{Table of Contents}
\renewcommand{\cfttoctitlefont}{
\hrule
\vspace{10pt}
\Large \bfseries}

\cftsetindents{section}{0.5in}{0.5in}
\renewcommand{\cftaftertoctitle}{
\vspace{10pt}%
\hrule
\vspace{36pt}}
\tableofcontents

\newpage

\renewcommand{\thesection}{\large{S.\arabic{section}}}
\section{\large Verifying the Polarization State of the Synthesized Beam}
As discussed in the main context, the synthesized field, which is constructed by individual optical emitters with different orientations specified by their azimuthal angle $\theta$, can be written as
\renewcommand{\theequation}{S\arabic{equation}}
\setcounter {equation} {0} 
\begin{equation}
\vec{E}\sim\begin{pmatrix} -\sin\theta \\ \cos\theta \end{pmatrix}\cdot e^{i\cdot l\cdot\theta}=\frac{i}{2}\begin{pmatrix} 1\\ -i \end{pmatrix}\cdot e^{i\cdot (l+1)\cdot\theta}-\frac{i}{2}\begin{pmatrix} 1\\ i \end{pmatrix}\cdot e^{i\cdot (l-1)\cdot\theta}
\end{equation}
where the first term on the right-hand side is a right-hand circularly polarized (RHCP) beam with OAM state $l+1$ while the second term is a left-hand circularly polarized (LHCP) beam with OAM state $l-1$. After passing through a quarter-wave plate whose fast axis is aligned in the vertical direction, the resulting RHCP component in the synthesized beam becomes
\begin{equation}
E_{RHCP}=\begin{pmatrix} 1 & 0 \\ 0 & -i \end{pmatrix}\cdot \frac{i}{2} \begin{pmatrix} 1 \\ -i \end{pmatrix}\cdot e^{i\cdot (l+1)\cdot\theta}=\frac{i}{2}\begin{pmatrix} 1\\ -1 \end{pmatrix}\cdot e^{i\cdot (l+1)\cdot\theta}
\end{equation}
which is a linearly polarized beam aligned 45$^\circ$ to the fast axis with OAM state $l+1$, as shown in Fig. S1. Similarly, the LHCP component in the synthesized beam is turned into 
\begin{equation}
E_{RHCP}=\begin{pmatrix} 1 & 0 \\ 0 & -i \end{pmatrix}\cdot (-\frac{i}{2}) \begin{pmatrix} 1 \\ i \end{pmatrix}\cdot e^{i\cdot (l-1)\cdot\theta}=-\frac{i}{2}\begin{pmatrix} 1\\ 1 \end{pmatrix}\cdot e^{i\cdot (l-1)\cdot\theta}
\end{equation}
which is a linearly polarized beam aligned -45$^\circ$(or 135$^\circ$) to the fast axis with OAM state $l-1$, as shown in Fig. S1. Here the minus sign is because the angle is measured counterclockwise by convention. 

Therefore, the RHCP component and the LHCP component in the synthesized beam can be distinguished and observed by inserting an additional linear polarizer with appropriate alignment of its axis to that of the quarter-wave plate, as shown in Fig. S1. For example, when the transmission axis of the linear polarizer is placed 45$^\circ$ to the fast axis of the quarter-wave plate, the RHCP light is projected into and viewed by the infrared camera; while the LHCP light is imaged when the polarizer is rotated -45$^\circ$ to the fast axis.

\clearpage
\begin{figure}
\centerline{\includegraphics [width = 16cm] {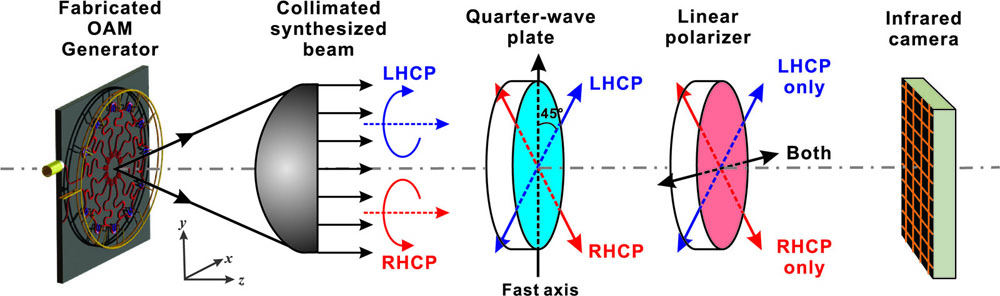}}
\end{figure}
\noindent {\bf Fig. S1.} \textbf{Experimental setup to verify the polarization state of the synthesized beam.} The synthesized field was first collimated by an objective lens and was then turned into two orthogonal linear polarizations by a quarter-wave plate. Each polarization was then imaged by an infrared camera after filtering out the other polarization. 

\clearpage
\section{\large Identifying the OAM State of the Synthesized Beam}
As shown in the inset of Fig. S2, when the OAM beam interferes with a co-propagating Gaussian reference beam, a spiral-like interference pattern can be generated. The $l$-value can hence be identified by the number of arms in the spiral pattern, and the sign of $l$ can be seen from the chirality or handedness of the spiral. This provides a convenient way to experimentally determine the OAM state of the synthesized beam. As shown in Fig. S2, a laser input was first divided into two arms using a 10:90 fiber coupler. The lower arm was then coupled into the silicon photonic chip after passing through a polarization controller (PC2) and a variable optical attenuator (VOA). PC2 was adjusted so that light launched into the photonic circuit was predominantly TE-polarized which is the operating polarization on-chip. The OAM beam was then generated from the chip and collimated by the objective lens. Similarly, light in the upper arm also passed through a polarization controller (PC2) and a VOA before it emitted from the end of the fiber to generate a Gaussian reference beam. The OAM beam and the Gaussian beam were then combined by an optical beam splitter and projected into the infrared camera. The polarization state of the reference arm was adjusted to RHCP using PC1 to interfere with the RHCP component of the OAM beam. In addition, the VOAs in the two interference arms were adjusted accordingly to balance the intensity of the two beams to generate an ideal interference pattern in the infrared camera.  

\clearpage
\begin{figure}
\centerline{\includegraphics [width = 16cm] {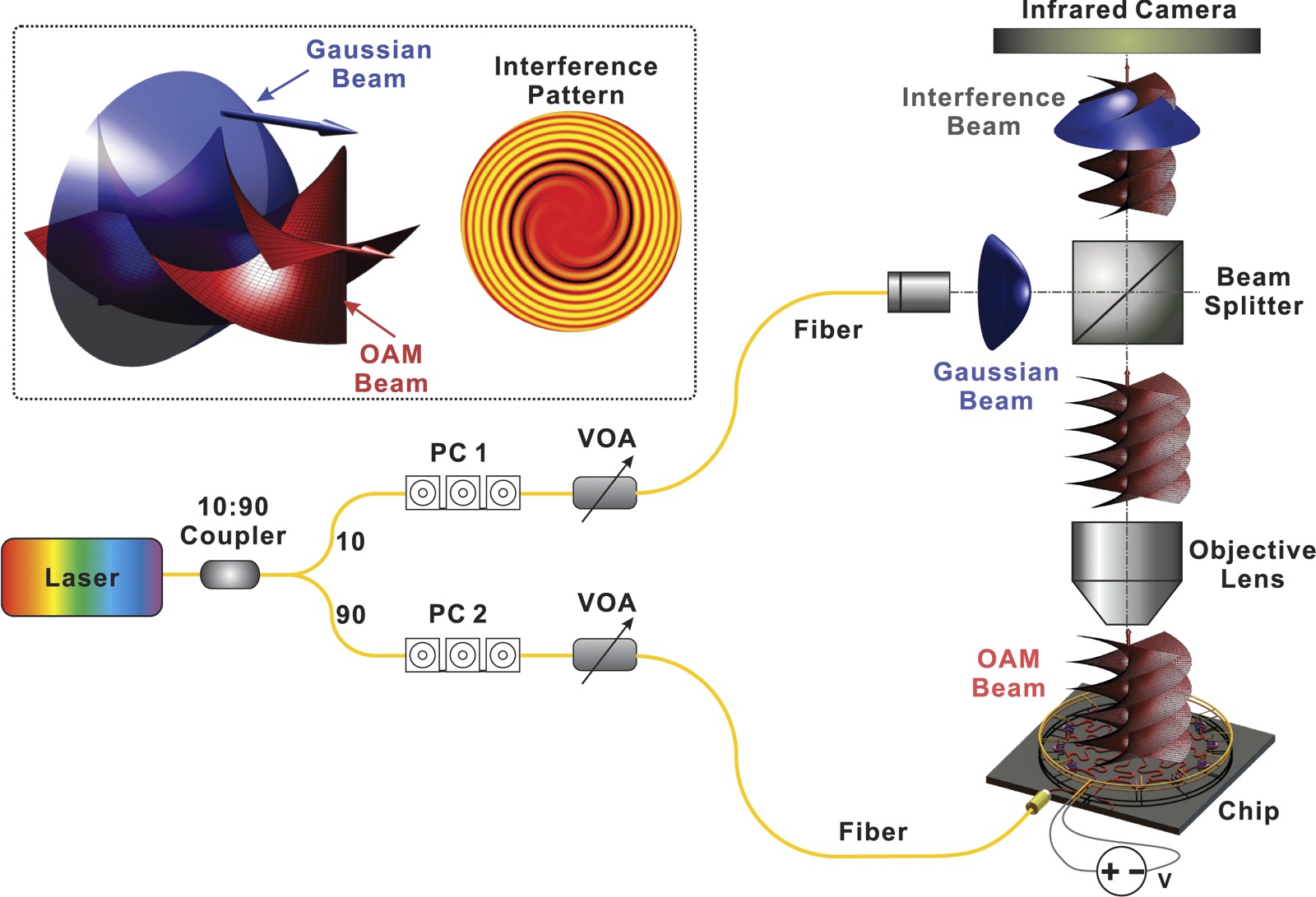}}
\end{figure}
\noindent {\bf Fig. S2.} \textbf{Experimental setup to identify the OAM state of the synthesized beam.} The inset illustrates the interference of the OAM beam with a Gaussian reference beam to reveal the helical phase front of the OAM beam from which the $l$-value can be identified. PC: Polarization Controller, VOA: Variable Optical Attenuator. 

\clearpage
\section{\large Other Supplementary Materials}
\subsection*{\normalsize Movie S1}
This movie simulated the evolution of the spiral interference pattern of a Gaussian reference beam with a co-propagating OAM beam generated from the proposed device, where the order of the OAM beam continuously changed from -4 to +4. This reveals the transition of an OAM beam from one integer state to the next as well as a continuum of OAM states inbetween two consecutive integer OAM states. 

\vspace{10mm}
\subsection*{\normalsize Movie S2}
This movie experimentally demonstrated the evolution of the spiral interference pattern of a Gaussian reference beam with an OAM beam generated from the fabricated device, where the order of the OAM beam continuously changed from -4 to +4 by adjusting the voltage applied on the phase shifters. This experimentally verifies the continuous tunability of the fabricated OAM generator with a single electrical signal and also reveals the continuous transition of an OAM beam from one integer state to the next as well as a continuum of OAM states inbetween two consecutive integer OAM states. The measurement agrees well with the simulation shown in Movie S1. Note that the vibration in the measured interference pattern was caused by the relative phase fluctuation in the Gaussian reference arm, which was induced by the mechanical perturbations of the interferometer, the fiber, and the environment, and which does not represent the stability of the OAM beam.   
\end{document}